\documentclass[twocolumn,showpacs,preprintnumbers,amsmath,amssymb]{revtex4}
\usepackage{graphicx}
\usepackage{dcolumn}
\usepackage{bm}
\begin{document}
\title{Static dielectric function with exact exchange 
contribution in the electron liquid}
\author {Zhixin Qian}
\affiliation{Department of Physics,
Peking University, Beijing 100871, China}
\date{\today}
\begin{abstract}
The exchange contribution, $\Pi_1 ({\bf k}, 0)$, to the static dielectric 
function in the electron liquid is evaluated exactly.
Expression for it is derived analytically in terms of one quadrature. 
The expression, as presented in Eq. (3) in the Introduction, turns
out to be very simple. A fully explicit expression (with no more 
integral in it) for $\Pi_1 ({\bf k}, 0)$ is further developed in terms of 
series.
Equation (3) is proved to be equal to the expression obtained
before under some mathematical assumption by Engel and Vosko, thus 
in the meanwhile putting the latter on a rigorous basis.
The expansions of $\Pi_1 ({\bf k}, 0)$ at 
the wavectors of $k=0$, $k=2k_F$, and at limiting large $k$ are derived.
The results all verify those obtained 
by Engel and Vosko.
\end{abstract}

\maketitle

\section{Introduction with concluding remarks}

The dielectric function in the homogeneous electron
liquid is determined by \cite{Fetter,Pines}

\begin{equation}
\epsilon ({\bf k} , \omega) = 1 - v(k) 
\Pi ({\bf k}, \omega) ;
\end{equation}
$\Pi ({\bf k}, \omega)$ is the proper (irreducible) linear response
function. [$v(k)$ is the Fourier transform of the Coulomb potential.]
Many electronic properties in solids and liquids can be 
described from the dielectric function. Rapid developments in 
the techniques of inelastic x-ray scattering 
and electron energy-loss spectroscopies
in recent years \cite{Burns,Waidmann,Schulke,Huotari,Glenzer}
now enable us to measure these dielectric function related properties 
with unprecedented accuracy. 

Theoretically,
Lindhard \cite{Lindhard} obtained, about six decades ago,
the explicit expression for $\Pi_0 ({\bf k}, \omega)$,
the zeroth-order of $\Pi ({\bf k}, \omega)$ in terms of 
Coulomb interaction, in the homogeneous electron gas; 
wherefore it is also known as Lindhard function. 
Keeping $\Pi_0 ({\bf k}, \omega)$ only in Eq. (1) yields
the random phase approximation (RPA) \cite{Fetter,RPA}
for $\epsilon ({\bf k} , \omega)$.
An enormous amount of effort had been devoted to
the study of the correction beyond RPA since the pioneering
proposals made by Hubbard \cite{Hubbard} (and 
DuBois \cite{DuBois}).
The correction beyond RPA
is fully due to the exchange-correlation effects,
which is also termed, according to Hubbard, as local field correction
$G({\bf k}, \omega)$:
\begin{equation}
G({\bf k}, \omega) =v(k)^{-1} \biggl [ 
\frac{1}{\Pi ({\bf k}, \omega)} 
-\frac{1}{\Pi_0 ({\bf k}, \omega)} \biggr ].
\end{equation}

Needless to say, the lowest order correction,
$\Pi_1 ({\bf k}, \omega)$, to the Lindhard function,
which is traditionally also known as the exchange contribution to
$\Pi ({\bf k}, \omega)$, had been of major research interest
\cite{Geldart1,Toigo,Rasolt,Sham,Tripathy,Rao,Brosens,Holas,Alvarellos,Kleinman,Chevary}.
In some of the researches, not strictly $\Pi_1 ({\bf k}, \omega)$
but variants of it were studied as well. Higher order contributions beyond 
$\Pi_0 ({\bf k}, \omega)$$+\Pi_1 ({\bf k}, \omega)$ had also been studied, 
for instance, in Refs. \cite{Geldart1,Holas}. 
About two and a half decades ago, Engel and Vosko \cite{Engel} obtained
an analytical expression [Eq. (29) in Ref. \cite{Engel}] for 
$\Pi_1 ({\bf k}, 0)$ in terms of
one quadrature. That was definite progress, and one must be 
aware that the original form for $\Pi_1 ({\bf k}, 0)$ is in
six-fold integral [see Eq. (\ref{Def-pi}) in the next section]. 
Indeed, most of the work we mentioned above relied on numerical procedures
in one way or another. 
Unfortunately the derivation by Engel and Vosko can not be fully regarded as rigorous
in that it critically relies on some mathematical assumption.
But convergence to the numerical results for $\Pi_1 ({\bf k}, 0)$
and many right analytical features all strongly suggest the correctness
of their expression. Very
encouragingly in this connection, Glasser \cite{Glasser}  
confirmed Eq. (29) of Ref. \cite{Engel} with the aid of analytical 
computer programs. Several of the leading terms in the small
wavevector expansion of $\Pi_1 ({\bf k}, 0)$ obtained from it were also
confirmed later analytically by Svendsen and von Barth \cite{Svendsen}.

In this paper we report an exact evaluation of $\Pi_1 ({\bf k}, 0)$.
The main result of the present work is the following
expression for $\Pi_1 ({\bf k}, 0)$,

\begin{widetext}
\begin{equation}  \label{intro}
\Pi_1 ({\bf k}, 0)= \frac{m^2e^2}{2\pi^3 k^2}  \biggl [
b \ln^2  \biggl | \frac{b}{a} \biggr | \biggl (\frac{1}{3}
a \ln \biggl |\frac{b}{a} \biggr | + b \biggr )
+\int_{a^2}^{b^2}
dx \frac{1}{x - a^2} \ln \biggl |\frac{x}{a^2} \biggr | 
\biggl (\frac{1}{2}ab \ln \biggl |\frac{x}{b^2} \biggr | -k \biggr )
\biggr ] ,
\end{equation}
\end{widetext}
with $a=1-k/2$, $b=1+k/2$, and $k$ in units of $k_F$. 
The result for $\Pi_1 ({\bf k}, 0)$ can be claimed to be 
surprisingly simple.
Equation (\ref{intro}) is proved to be equal to
the one given by Engel and Vosko \cite{Engel} mentioned above; 
but in the present work it is derived rigorously.
The expansions of it at
$k=0$, $k=2$, and at limiting large $k$ are also derived; they
fully confirm the corresponding ones obtained in Ref. \cite{Engel}.

The remaining integral in Eq. (\ref{intro}) is a straightforward 
numerical task. But we have an interesting 
alternative which is to further 
carry out {\em analytically} the remaining integral in terms of series.
In this way, a fully explicit expression for $\Pi_1 ({\bf k}, 0)$ is 
developed, which is shown in Eq. (\ref{series}) in Sec. V, with the 
quantity I(k) being related to $\Pi_1 ({\bf k}, 0)$ as in 
Eq. (\ref{Def-I)}), and $\zeta(3)$ being Riemann's zeta function. 
This series representation  
should be suitable for both of numerical and analytical applications.

The details of our derivation are given in Sections II-IV; 
supporting materials are relegated to Appendices A and B.
The final result is given in Sec. V in which the above-mentioned
series representation is also developed. 
The proof of the equivalence of our result
and Eq. (29) in Ref. \cite{Engel} is given 
in Sec. VI.
The expansions of $\Pi_1 ({\bf k}, 0)$ 
in limiting cases are derived in Sec. VII.

\section{Reduction of $\Pi_1 ({\bf k}, 0)$ to two-dimensional integral}

Explicit expression for $\Pi_1 ({\bf k}, \omega)$ can be obtained with
the diagrammatic techniques \cite{DuBois} (see also Refs. 
\cite{Geldart1,Holas}): 
\begin{widetext}
\begin{equation}
\Pi_1 ({\bf k}, \omega) =2 
\int  \frac{d{\bf p}}{(2 \pi)^3} \frac{d{\bf p}'}{(2 \pi)^3} 
v({\bf p}- {\bf p}')
\frac{(n_{\bf p} -n_{{\bf p}+{\bf k}})(n_{{\bf p}'}
-n_{{\bf p}'+{\bf k}})}
{\omega +\omega_{\bf p} -\omega_{{\bf p}+{\bf k}} +i0^+}   
\biggl [ \frac{1}{\omega +\omega_{\bf p}
-\omega_{{\bf p}+{\bf k}} +i0^+} -\frac{1}{\omega +\omega_{{\bf p}'}
-\omega_{{\bf p}'+{\bf k}} +i0^+} \biggr ] ,
\end{equation}
where $\omega_{\bf p} =p^2/2m$
and $n_{\bf p}$ is the Fermi-Dirac distribution function. (We put $\hbar =1$
in this paper.)
At $\omega =0$, one gets
\begin{eqnarray}  \label{Def-pi}
\Pi_1 ({\bf k}, 0) = -8 \pi m^2e^2 \int \frac{d {\bf p}}{(2 \pi)^3} 
\int \frac{d {\bf p}'}{(2 \pi)^3} 
n_{{\bf p}-{\bf k}/2} n_{{\bf p}'-{\bf k}/2}    
\biggl [  \biggl (\frac{1}{{\bf p} \cdot {\bf k}} + 
\frac{1}{{\bf p}' \cdot {\bf k}} \biggr )^2 \frac{1}{|{\bf p} 
+ {\bf p}'|^2} - \biggl (\frac{1}{{\bf p} \cdot {\bf k}} -   
\frac{1}{{\bf p}' \cdot {\bf k}} \biggr )^2 
\frac{1}{|{\bf p} - {\bf p}'|^2}  \biggr ] .
\end{eqnarray}
\end{widetext}
The $\Pi_1 ({\bf k}, 0)$ in Eq. (\ref{Def-pi}) is equal 
to the corresponding one defined in Ref. \cite{Engel}.
We also define $I(k)$ as  
\begin{eqnarray}  \label{Def-I)}
I(k)=\frac{\pi^3}{m^2e^2} \Pi_1 ({\bf k}, 0) .
\end{eqnarray}
We note that $\Pi_1 ({\bf k}, 0)$ depends only on the magnitude $k$ 
in a uniform system. $I(k)$ so defined has the well-known property
of $I (k=0) =-1$ (see Sec. VII). 

We shall perform our calculation in the cylindrical
coordinates. The advantage of the cylindrical coordinates in calculating
$\Pi_1 ({\bf k}, \omega)$ had been appreciated before 
\cite{Brosens,Kleinman,Kleinman1,Engel,Glasser}.
The integrals over the azimuthal angular variables of
${\bf p}$ and ${\bf p}'$ can be readily carried out. After that 
one obtains
\begin{equation}  \label{I1}
I(k)=-\frac{1}{8k^2}  \int_{-a} \int^b dz dz'
\frac{1}{z^2 z'^2} [\alpha^2 L( \alpha^2 ) 
- \beta^2 L( \beta^2 ) ] ,    
\end{equation}
where $\alpha =z + z'$,  $\beta =z - z'$.
The same notations as those in Ref. \cite{Glasser} are adopted. 
The function $L$ in Eq. (\ref{I1}) is defined as
\begin{equation} \label{Def-L}
L(u) = \int_0^\lambda d  x
\int_0^ {\lambda '} d  x' \frac{1}{\sqrt{(x-x')^2 +2(x+x')u +u^2}} ,
\end{equation}
where $\lambda=(a+z)(b-z)$ and $\lambda '=(a+z')(b-z')$. 
The two-dimensional integral 
on the right hand side of Eq. (\ref{Def-L}) can be further 
carried through. The resulting  expression for $L(u)$ is
\begin{eqnarray} \label{L}
L(u) &=&  \frac{1}{2} [L_1 (u) -(\lambda +\lambda ')
(1 +2 \ln |2u|)- u]   \nonumber \\
&+& \lambda \ln | L_1 (u) + \lambda ' -\lambda +u |  \nonumber \\ 
&+& \lambda ' \ln | L_1 (u) + \lambda -\lambda ' +u | , 
\end{eqnarray} 
where
\begin{eqnarray} \label{L1}
L_1 (u) &=& \sqrt{(\lambda ' -\lambda +u)^2 +4 \lambda u}  \nonumber \\
&=& \sqrt{u^2 +(\lambda-\lambda ')^2 + 2(\lambda +\lambda ') u} .
\end{eqnarray}
Substituting Eq. (\ref{L}) into Eq. (\ref{I1}), one obtains
\begin{equation}  
I(k)=-\frac{1}{16k^2}  \sum_{i=0}^{3} J_i ,
\end{equation} 
where
\begin{equation}  \label{Def-J0}
J_0 = -4\int_{-a}  \int^b d z d z' \frac{1}{z z'}
[(\lambda +\lambda')(2\ln 2 +1)+\alpha^2 +\beta^2] ;
\end{equation}
\begin{eqnarray}  \label{Def-J1}
J_1 = \int_{-a}  \int^b d z d z' \frac{1}{z^2 z'^2}
[\alpha^2 L_1 ( \alpha^2 ) - \beta^2 L_1 ( \beta^2 ) ]  ; 
\end{eqnarray}
\begin{widetext}
\begin{eqnarray}  
J_2 = 2\int_{-a}  \int^b  dz d z' \frac{1}{z^2 z'^2} &\{& \lambda
[\alpha^2 \ln |L_1 (\alpha^2 ) +\lambda' -\lambda + \alpha^2 |
 - \beta^2 \ln |L_1 (\beta^2 ) +\lambda' -\lambda + \beta^2 | ] 
\nonumber \\
&+&\lambda'
[\alpha^2 \ln |L_1 (\alpha^2 ) +\lambda -\lambda' + \alpha^2 |
 - \beta^2 \ln |L_1 (\beta^2 ) +\lambda -\lambda' + \beta^2 | ] \},
\end{eqnarray}
or, by the use of the fact that both of $L_1(\alpha^2)$ and
$L_1(\beta^2)$ remain unchanged [in virtue of
the second identity in Eq. (\ref{L1})] upon exchanging $z$ and $z'$,
\begin{equation}  \label{Def-J2}
J_2 =4 \int_{-a}  \int^b  dz d z' \frac{1}{z^2 z'^2} \lambda
[\alpha^2 \ln |L_1 (\alpha^2 ) +\lambda' -\lambda + \alpha^2 | 
 - \beta^2 \ln |L_1 (\beta^2 ) +\lambda' -\lambda + \beta^2 | ] ;
\end{equation}
\end{widetext}
and
\begin{equation}  \label{Def-J3}
J_3 =4 \int_{-a}  \int^b d z d z' \frac{1}{z^2 z'^2} \lambda
[\beta^2 \ln |\beta ^2 | - \alpha^2 \ln | \alpha^2  | ] . 
\end{equation}
$J_0$ can be evaluated to be
\begin{equation}  
J_0=-8 \ln|b/a| [ ab(1+2 \ln2) \ln|b/a| + k (3 +\ln 2)],
\end{equation}
and accordingly
\begin{eqnarray}  \label{I-123}
I(k)=-\frac{1}{16k^2}                                 
\{ & - &8 \ln |b/a| [ab(1+2 \ln 2) \ln |b/a|   \nonumber \\
& + & k(3 +2 \ln 2 ) ] +\sum_{i=1}^{3} J_i \} .
\end{eqnarray}
The $J_1$, $J_2$, and $J_3$ are the same as the ones defined
in Ref. \cite{Glasser}. The expression for $J_1$ here appears to 
be different to that in Ref. \cite{Glasser} but is actually equal.
There is an extra term of the following form,
\begin{equation}  \label{proof0}
\int_{-a} d z \int^b d z' \frac{1}{z^2 z'^2} (\lambda'  -\lambda )
[L_1 ( \alpha^2 ) -  L_1 ( \beta^2 ) ] ,
\end{equation}
in the expression for $J_1$ in Eq. (10) of Ref. \cite{Glasser},
but this extra term can be shown to be equal to zero. 
(We note that in our foregoing derivation this term 
did not arise at all.)
In fact, the factor of $\lambda  -\lambda '$ changes sign but that of 
$L_1 ( \alpha^2 ) -  L_1 ( \beta^2 )$ [actually each
of $L_1 ( \alpha^2 )$ and $L_1 ( \beta^2 )$] remains unchanged
upon exchanging $z$ and $z'$ as mentioned above. 
In consequence, 
\begin{equation}  \label{proof}
\int_{-a} d z \int^b d z' \frac{1}{z^2 z'^2} (\lambda'  -\lambda )
[L_1 ( \alpha^2 ) -  L_1 ( \beta^2 ) ] = 0 . 
\end{equation}
Understandably, this term with null contribution might 
however cost a huge amount of labor in any further numerical or analytical 
calculations if not appreciated.

The expression for $I(k)$ has as a consequence been successfuly
reduced in terms of two-dimensional integral. Similar type of reduction
had been accomplished or applied in 
Refs. \cite{Brosens,Kleinman,Kleinman1,Chevary,Engel,Glasser}.  
(Brosens et al. \cite{Brosens} in fact
considered both of the static and dynamic cases.) A different
approach was adopted by Holas et al. \cite{Holas} who essentially first 
calculated the imaginary part of $\Pi_1 ({\bf k}, \omega)$,
and then the real part of it [with the static
$\Pi_1 ({\bf k}, 0)$ as its special case]
via the dispersion relation, which virtually
also amounts to a two-dimensional integral over all.   
The remaining two-dimensional integral was then
calculated numerically in 
Refs. \cite{Brosens,Holas,Kleinman1,Kleinman,Chevary}; analytically with 
the aid of some ingenious method and mathematical assumption
in Ref. \cite{Engel}; and analytically but with computer programs
in Ref. \cite{Glasser}.
We need point out that there exists discrepance between 
our Eq. (\ref{I-123}) and the corresponding 
expression in Ref. \cite{Glasser} [the one shown in the
middle of the paragraph above
Eq. (10) in Ref. \cite{Glasser}]. 

\section{Evaluation for $J_1$}

It turns out that the integral in Eq. (\ref{Def-J1}) 
for $J_1$ can be \emph{fully} carried through.
We are about to achieve this step by step. Indeed, either the integral 
over $z$ or $z'$ can be carried out first. But before doing that 
it is worth pointing out 
a somewhat hidden but useful truth. $L_1(\beta^2 )$ can actually be, 
by direct substitution in
Eq. (\ref{L1}), 
simplified to the following form:
\begin{eqnarray}  \label{L1-beta}
L_1 ( \beta^2 )  =  2 |\beta| .
\end{eqnarray}
Accordingly $J_1$, with some further algebra, can be written as
\begin{eqnarray}  
J_1 = 4 \int_{-a}^b dz \frac{1}{z^2} \int_{-a}^b d z' \frac{1}{z'}
[\alpha \sqrt{R(z, z')} + \beta |\beta | ] ,
\end{eqnarray}
where 
\begin{eqnarray}  \label{R}
R(z, z') = C_0 (z) z'^2 + B_0 (z) z' +A_0 (z) ,
\end{eqnarray}
with
\begin{equation}  \label{ABC}
A_0 (z) = z^2, ~ B_0 (z) = (2+2kz -k^2)z,~ 
C_0 (z) = 1+ 2kz .  
\end{equation} 
For brevity, $A_0 (z)$, $B_0 (z)$, and $C_0 (z)$ will be
denoted, respectively, as $A_0 $, $B_0 $, and $C_0 $ instead, 
i.e., with the argument $z$ suppressed. 
We next define 
\begin{eqnarray}  \label{Def-J1-A}
{\bar J}_1^A (z) =\int_{-a}^b d z' \frac{1}{z'}\alpha \sqrt{R(z, z')} ,
\end{eqnarray}
and 
\begin{eqnarray}  \label{Equ0}
{\bar J}_1^B (z) =\int_{-a}^b d z' \frac{1}{z'} \beta |\beta| ,
\end{eqnarray}
so that we can write $J_1$ as
\begin{eqnarray}  \label{J1-next}
J_1 = 4 \int_{-a}^b dz \frac{1}{z^2} [ {\bar J}_1^A (z) 
+ {\bar J}_1^B (z) ] .
\end{eqnarray}
The integration over $z'$ on the right hand side of Eq. (\ref{Equ0})
for ${\bar J}_1^B (z)$ presents no analytical difficulty. 
It is straightforward and the resulting 
expression is,
\begin{equation} \label{J1-B}
{\bar J}_1^B (z) = 2kz -1 - k^2/4 -z^2 (3- 2 \ln |z| + \ln |ab|).  
\end{equation}
On the other hand, the integration in ${\bar J}_1^A (z)$ is more subtle,
but becomes also routine \cite{Gradshteyn} if we rewrite it as
\begin{equation}  \label{manu1}
{\bar J}_1^A (z) = z \int_{-a}^b d z' \frac{1}{z'} \sqrt{R(z, z')} 
+ \int_{-a}^b d z' \sqrt{R(z, z')} .
\end{equation}
Each of the integrations in Eq. (\ref{manu1}) can be carried through, 
and the result for ${\bar J}_1^A (z)$ is
\begin{eqnarray}  \label{J1-A}
{\bar J}_1^A (z) &=& 1  + \frac{1}{4}k^2 +\frac{5}{2}kz +2z^2  
 + \frac{B_0}{4C_0}  (2z +k)         \nonumber \\
 &-& z |z| \ln \biggl |\frac{a \psi_1(z)}
{b \psi_2(z)} \biggr |   \nonumber \\
&+& \frac{1}{2} \bigg ( zB_0 + A_0 - \frac{B_0^2}{4C_0} \biggr ) 
\frac{1}{\sqrt{C_0}}
\ln \biggl |\frac{ \psi_3(z)}{ \psi_4(z)} \biggr | ,
\end{eqnarray}
where 
\begin{eqnarray}  \label{psi1}
\psi_1(z) = 2A_0 +B_0 b + 2 \sqrt{A_0 R(z, b)} ,
\end{eqnarray}
\begin{eqnarray}   \label{psi2}
\psi_2(z) = 2A_0 - B_0 a + 2 \sqrt{A_0 R(z, -a)} ,
\end{eqnarray}
\begin{eqnarray}    \label{psi3}
\psi_3(z) =  2 \sqrt{C_0 R(z, b)} +2C_0 b +B_0 ,
\end{eqnarray}
and
\begin{eqnarray}    \label{psi4}
\psi_4(z) =  2 \sqrt{C_0 R(z, -a)} -2C_0 a +B_0 .
\end{eqnarray}
$R(z, b)$ and $R(z, -a)$ can be written explicitly in the following forms,
\begin{eqnarray}
R(z, b) = [(1+k)z +b]^2 ; ~
R(z, -a) = [(k-1)z +a]^2 .   \nonumber \\
\end{eqnarray}
Equation (\ref{J1-next}), together with the explicit expressions 
(\ref{J1-B}) for ${\bar J}_1^B (z)$ and  (\ref{J1-A}) 
for ${\bar J}_1^A (z)$, indicates that we have already reduced $J_1$ in terms
of one quadrature. Further analytical refinement turns out to be 
practicable, but only
after one finds a way to simplify the somewhat unwieldy quantities 
$\psi_1(z) /\psi_2(z)$ and $\psi_3(z) /\psi_4(z)$.
Indeed, we find the following simple expressions for them: 
\begin{eqnarray} \label{psi1/psi2}
\frac{\psi_1(z)}{\psi_2(z)} = \biggl (\frac{2b}{k} \biggr )^2 \theta (z)
+  \biggl (\frac{k}{2a} \biggr )^2 \theta (-z) ,
\end{eqnarray} 
where $\theta (z) =1$ for $z >0$ and $\theta (z) =0$ for $z <0$; and
\begin{eqnarray} \label{psi3/psi4}
\frac{\psi_3(z)}{\psi_4(z)} = 
\biggl (\frac{\sqrt{C_0} +1}{\sqrt{C_0} -1} \biggr ) ^2 .
\end{eqnarray}
The simplifications play a critical role in our further 
reducing $J_1$ to a fully explicit form [Eq. (\ref{Fin-J1}) below]. 
The verification for Eq. (\ref{psi1/psi2}) 
and Eq. (\ref{psi3/psi4}) is given in Appendix A. We
here substitute them into Eq. (\ref{J1-A}) to obtain
\begin{eqnarray} \label{Fin-J1-A}
{\bar J}_1^A (z) &=& 1 + \frac{1}{4}k^2 +\frac{5}{2}kz
+ \biggl ( 2 - \ln \biggl |1 -\frac{4}{k^2} \biggr |
\biggr ) z^2
\nonumber \\
&+& \frac{B_0}{4C_0}  (2z +k)  \nonumber \\
&+& \frac{1}{4C_0^{3/2}} z^2 [8 -k^4
+ 4kz (6 -k^2) + 12k^2 z^2 ] Y(z) ,   \nonumber \\
\end{eqnarray}
where we have adopted the following notation,
\begin{eqnarray} 
Y(z) = \ln \biggl | \frac{\sqrt{C_0} +1}{\sqrt{C_0} -1} \biggr | .
\end{eqnarray}
We then substitute 
Eqs. (\ref{J1-B}) and (\ref{Fin-J1-A}) into 
Eq. (\ref{J1-next}) and, with some
further routine calculation, obtain
\begin{eqnarray} \label{equation1}
J_1 = &-& 20 +16 \ln k + \frac{2}{k} (k^2 -1)^2 \ln 
\biggl | \frac{k+1}{k-1} \biggr |  \nonumber \\ 
&-& (k^3 -24 k +8) \ln|2b|   \nonumber \\
&+& (k^3 -24 k -8) \ln|2a|    \nonumber \\
&+& \frac{1}{k}[3 \eta_1 + 2(3-k^2) \eta_{-1}-(k^2 -1 )^2 \eta_{-3}] ,
\end{eqnarray}
where
\begin{eqnarray}   \label{eta-n}
\eta_n = k \int_{-a}^b dz C_0^{n/2} Y(z) .
\end{eqnarray}
The explicit expressions for $\eta_1$, $\eta_{-1}$, and $\eta_{-3}$ are
given in Eq. (\ref{I1/2}), Eq. (\ref{I-1/2}), and Eq. (\ref{I-3/2}),
respectively. 
By substituting Eqs. (\ref{I1/2}), (\ref{I-1/2}), (\ref{I-3/2}) 
into Eq. (\ref{equation1}), we obtain our final result for $J_1$: 
\begin{eqnarray} \label{Fin-J1}
J_1=-16 + 4 (5k + 4/k) \ln |b/a| .
\end{eqnarray}
The result is surprisingly elegant.

\section{Evaluation for $J_2 +J_3$}

Among the three expressions as shown in (\ref{Def-J1}), 
(\ref{Def-J2}), and (\ref{Def-J3}), the expression in (\ref{Def-J1}) 
for $J_1$ is likely the one the most amenable 
to analytical computations. 
This is in fact also the cause that it can be fully reduced 
to the explicit expression of (\ref{Fin-J1}). On the other hand, 
due to the logarithm forms
in them, Eq. (\ref{Def-J2}) and Eq. (\ref{Def-J3}) are much more difficult.
Further reduction of $J_2$ and $J_3$, at least that of $J_2$ , to
one quadrature, had been believed beyond barehanded
effort \cite{Glasser}. We find indeed that careless choices 
of procedures might easily make the task intractable. It is critical 
to search for optimal ones.

Instead of evaluating $J_2$ and $J_3$ separately, we choose to combine
them together and deal with them at one stroke. 
We write $J_{23} $$=J_2$$+J_3$, and accordingly have
\begin{widetext}
\begin{equation}  \label{Def-J23}
J_{23} =4 \int_{-a}  \int^b dz d z' \frac{\lambda}{z^2 z'^2}    
 \left\{\alpha^2 \ln | [L_1 (\alpha^2 )
+\lambda' -\lambda +\alpha^2]/ \alpha^2 |
- \beta^2 \ln |[a-b +2z +2\beta/|\beta|] /
\beta  | \right\} .
\end{equation}
We have also made the use of Eq. (\ref{L1-beta}) in obtaining the above
equation.
We next perform some manipulation on Eq. (\ref{Def-J23}) to bring it
into the following form:
\begin{equation}  \label{J23-next}
J_{23}=16 \int_{-a}^b d z 
\frac{1}{z}\lambda \ln|4 \lambda |\int_{-a}^b d z' 
\frac{1}{ z'}    - 4 N ,
\end{equation}
where
\begin{equation}  \label{Def-N}
N = \int_{-a} \int^b  dz d z' \frac{\lambda}{z^2 z'^2} 
[\alpha^2 \ln | \alpha^2 +\lambda' -\lambda - 2\sqrt{R(z, z')} |  
- \beta^2 \ln | \beta (a-b + 2z - 2\beta /|\beta|  ) | ] .
\end{equation}
\end{widetext}
The first term on the right hand side of Eq. (\ref{J23-next}) has 
virtually been reduced to a one-dimensional integral for the integration
$\int_{-a}^b dz' 1/z'$ is trivial.
Hence we need concern ourselves only with the term of $-4N$. 
For this purpose, we first rewrite $N$ as the following:
\begin{eqnarray}  \label{N}
N = \int_{-a}^b d z \frac{ \lambda}{z^2}[{\bar N}_1(z) -{\bar N}_2(z)] ,
\end{eqnarray}
where 
\begin{eqnarray}  \label{Def-N1}
{\bar N}_1(z) = \int_{-a}^b d z' \frac{ \alpha^2}{z'^2} 
\ln | \alpha^2 +\lambda' -\lambda - 2\sqrt{R(z, z')} | ,
\end{eqnarray}
and
\begin{equation}  \label{Def-N2}
{\bar N}_2(z) = \int_{-a}^z d z' \frac{ \beta^2}{z'^2}
\ln |2 \beta (z-b) | + \int_{z}^b d z' \frac{ \beta^2}{z'^2}
\ln |2 \beta (z+a) | .  
\end{equation}
In the preceding expression for ${\bar N}_2 $, the integral over $z'$ 
has been purposely separated into two regimes of $-a \le z' < z$ and
$z \le z' \le b$. This is intended to take care of the
subtle singular behavior of the logarithm factor 
$\ln | \beta (a-b + 2z - 2\beta /|\beta|) |$    
in Eq. (\ref{Def-N}).
After performing partial integration on each of the integrals
on the right hand sides of
Eq. (\ref{Def-N1}) and Eq. (\ref{Def-N2}),  one obtains
\begin{eqnarray}  \label{N1}
{\bar N}_1 (z) = &2&  \biggl (
1 - \frac{z^2}{ab} 
+ z \ln \biggl |\frac{b}{a} \biggr| \biggr ) 
\ln |2 \lambda |     \nonumber \\  
&+& \int_{-a}^b 
d z' \biggl ( z' 
- \frac{z^2}{z'} + 2z \ln |z'| \biggr )   \nonumber \\
&~&\times \frac{1}{\alpha} 
\biggl (\frac{kz}{\sqrt{R(z, z')}} - 1 \biggr ) ,
\end{eqnarray}
and
\begin{eqnarray}  \label{N2} 
{\bar N}_2(z) =&2&  \left(1 - \frac{z^2}{ab}  
- z \ln \biggl |\frac{b}{a} \biggr| \right)
\ln |2 \lambda |  + 2z W_1 (z)\ln |z|   \nonumber \\
&+& \int_{-a}^b d z'
\left(z' -\frac{z^2}{z'} - 2z \ln |z'|
\right) \frac{1}{\beta} .
\end{eqnarray}
Here we have introduced the following notation,
\begin{eqnarray}  \label{W1} 
W_1 (z) = \ln \biggl | \frac{z+a}{z-b} \biggr | . 
\end{eqnarray}
Equation (\ref{N}), together with 
Eqs. (\ref{N1}) and (\ref{N2}), 
is then substituted into Eq. (\ref{J23-next}) to get
\begin{eqnarray} \label{J23}
J_{23} = && 8 (2 \ln 2 -1)(ab \ln|b/a| +k) \ln |b/a|    \nonumber \\
&& +4 (-k \Phi_1 +2 \Phi_2 -2k  \Phi_3 +2\Phi_4) ,
\end{eqnarray}
where
\begin{eqnarray} \label{Def-Phi2}
\Phi_1= -\int_{-a}^b dz \frac{\lambda}{z} \int_{-a}^b dz'
\frac{\beta}{z'} \frac{1}{\sqrt{R(z, z')}} ,
\end{eqnarray}
\begin{eqnarray} \label{Def-Phi1}
\Phi_2= -2 \int_{-a}^b dz \frac{\lambda}{z} \int_{-a}^b dz'
\frac{z'}{\alpha \beta} \ln |z'| ,
\end{eqnarray}
\begin{eqnarray} \label{Def-Phi3}
\Phi_3=  \int_{-a}^b dz \lambda \int_{-a}^b dz'
\frac{1}{\alpha \sqrt{R(z, z')}} \ln |z'| ,
\end{eqnarray}
and
\begin{eqnarray} \label{Def-Phi4}
\Phi_4=  \int_{-a}^b dz \frac{\lambda}{z}  W_1 (z) \ln |z| .
\end{eqnarray}
$\Phi_4$ is clearly not our concern temporarily, since it is already 
in terms of one-dimensional integral. Fortunately it turns out that 
the two-dimensional integral in $\Phi_1$ can be fully carried through. 
We first attack it.
To this end, we write
\begin{eqnarray} \label{Phi2}
\Phi_1=  \int_{-a}^b dz \frac{\lambda}{z} {\bar \Phi}_1(z) ,
\end{eqnarray}
where
\begin{equation} \label{Def-Phi2bar}
{\bar \Phi}_1(z) =\int_{-a}^b dz' \frac{1}{\sqrt{R(z, z')}}
- z \int_{-a}^b dz' \frac{1}{z' \sqrt{R(z, z')}} .
\end{equation}
Each of the integrals over $z'$ in Eq. (\ref{Def-Phi2bar}) is routine.
After carrying out them, we obtain
the following result for ${\bar \Phi}_1(z) $,
\begin{eqnarray} 
{\bar \Phi}_1(z) = \frac{z}{\sqrt{A_0}}  \ln \biggl |
\frac{a \psi_1(z)}{b \psi_2(z)} \biggr |
+ \frac{1}{\sqrt{C_0}}\ln \biggl | \frac{\psi_3(z)}{\psi_4(z)} \biggr | ,
\end{eqnarray}
or, with the aid of Eqs. (\ref{psi1/psi2}) and (\ref{psi3/psi4}),
\begin{eqnarray} \label{Phi2bar}
{\bar \Phi}_1(z) = \frac{2}{\sqrt{C_0}}Y(z)  
+ \ln \biggl |\frac{4ab}{k^2} \biggr | .
\end{eqnarray}
Substitution of Eq. (\ref{Phi2bar}) into Eq. (\ref{Phi2}),
followed by further algebra, yields
\begin{eqnarray}  \label{Phi_1-man}
\Phi_1 = k(\ln |4ab| - 2 \ln k ) 
-\frac{1}{k^2} \eta_1 + \frac{1}{k^2}(1 +2k^2) \eta_{-1} .
\end{eqnarray}
The use of Eqs. (\ref{I1/2}) and (\ref{I-1/2}) in the 
above equation then yields 
\begin{eqnarray} \label{Fin-Phi2}
\Phi_1= && \frac{1}{3k^2} [(8k^3 +9k^2 +4 ) \ln b
+ (8k^3 -9k^2 -4 ) \ln |a|    \nonumber \\
&& - 4k +16 k^3 \ln|2/k| ] .
\end{eqnarray}
This is another elegant result indeed.

$\Phi_2$ in Eq. (\ref{Def-Phi1}) and $\Phi_3$ in 
Eq. (\ref{Def-Phi3}) are our next concern. Clearly one can do nothing
further directly with regard to integral over $z'$ in 
those expressions,
due to the simultaneous appearance
of the factor $\ln |z'|$ and nontrivial denominator
in each of the integrands. Our trick is to reverse the order of 
integrations over $z$ and over $z'$.
This means to perform the integration over $z$ first 
instead of that over $z'$.
This consideration yields the following form for $\Phi_2$,
\begin{eqnarray} 
\Phi_2=  2\int_{-a}^b dz z \ln |z| \int_{-a}^b dz' \frac{\lambda '}{z'}
 \frac{1}{\alpha \beta} .
\end{eqnarray}
Notice that we have exchanged the symbols of $z$ and $z'$
to obtain the preceding equation.
The integration over $z'$ can now be readily carried out.
After that one obtains
\begin{eqnarray} \label{Phi1}
\Phi_2=  \int_{-a}^b dz \frac{1}{z}\ln |z| 
&&[ 2ab \ln|b/a| + \lambda W_1 (z)    \nonumber  \\
&& -  (z^2 +kz -ab) W_2 (z) ] ,
\end{eqnarray}
where we have introduced the notation
\begin{eqnarray}    \label{W2}
W_2 (z) = \ln \biggl | \frac{z-a}{z+b} \biggr | .
\end{eqnarray}
This temporarily finishes our job for $\Phi_2$, for it is already
in terms of one quadrature.

Next we consider $\Phi_3$ of Eq. (\ref{Def-Phi3}) in the same fashion.
First notice the property 
\begin{eqnarray}
R(z, z') = R(z', z) ,
\end{eqnarray}
which enables us to rewrite $\Phi_3$ as
\begin{eqnarray} \label{equation2}
\Phi_3=  \int_{-a}^b dz  {\bar \Phi}_3 (z) \ln |z| ,
\end{eqnarray}
where
\begin{eqnarray} \label{equation3}
{\bar \Phi}_3(z)
=\int_{-a}^b dz' \frac{\lambda '}{\alpha \sqrt{R(z, z')}} .
\end{eqnarray}
Explicitly,
\begin{eqnarray}   \label{Equ1}
{\bar \Phi}_3(z) &=& (b+z)(a-z) \int_{-a}^b dz' 
\frac{1}{\alpha \sqrt {R(z, z')}}        \nonumber \\
&+& (k +z) \int_{-a}^b dz' \frac{1}{\sqrt{R(z, z')}}   
-\int_{-a}^b dz' \frac{z'}{\sqrt{R(z, z')}} .  \nonumber \\ 
\end{eqnarray}
Each of the integrals in Eq. (\ref{Equ1}) is routine.
They can be carried through and the result is 
\begin{eqnarray} \label{Phi3bar}
{\bar \Phi}_3(z)= &&  \frac{4 - (k+2z)^2}{4k|z|}
 \biggl [ \ln \biggl |\frac{\phi_1(z)}{\phi_2(z)} \biggr | 
- W_2 (z) \biggr ] -\frac{2z +k}{C_0}
             \nonumber   \\
&& + \frac{1}{2}(k+2z)(2 +3kz)  C_0^{-3/2} 
\ln \biggl |\frac{\phi_3(z)}{\phi_4(z)} \biggr | ,
\end{eqnarray}
where 
\begin{equation}  \label{phi1}
\phi_1(z) =2k^2z^2 -kz(2z+k)(z-a) +2k |z| \sqrt{R(z, -a)} ,
\end{equation}
\begin{equation}   \label{phi2}
\phi_2(z) =2k^2z^2 -kz(2z+k)(b+z) +2k |z| \sqrt{R(z, b)} ,
\end{equation}
\begin{equation}   \label{phi3}
\phi_3(z) =2 \sqrt{C_0 R(z, b)} +2(b+z)C_0 - kz(2z+k),
\end{equation}
and 
\begin{equation}  \label{phi4}
\phi_4(z) =2 \sqrt{C_0 R(z, -a)} +2(z-a)C_0 - kz(2z+k) .
\end{equation}
It is not difficult to verify that $\phi_3(z) = \psi_3(z)$ 
and $\phi_4(z) = \psi_4(z)$.
Therefore, according to Eq. (\ref{psi3/psi4}),
\begin{eqnarray} \label{phi3/phi4}
\frac{\phi_3(z)}{\phi_4(z)} =
\biggl ( \frac{\sqrt{C_0} + 1}{\sqrt{C_0} - 1} \biggr ) ^2 .
\end{eqnarray}
On the other hand, it is shown in Appendix A that
\begin{eqnarray} \label{phi1/phi2}
\frac{\phi_1(z)}{\phi_2(z)} =
 \theta (-z) \biggl ( \frac{z-a}{z+b} \biggr )^2 .
\end{eqnarray}
We next substitute Eqs. (\ref{phi3/phi4}) and (\ref{phi1/phi2}) 
into Eq. (\ref{Phi3bar}), and then the resultant equation further
into Eq. (\ref{equation2}). The result for $\Phi_3$ is in this way 
obtained as
\begin{eqnarray} \label{Fin-Phi3}
\Phi_3=  \int_{-a}^b dz && \ln |z|  \biggl [  -\frac{2z +k}{C_0}
- \frac{(a-z)(b+z)}{kz} W_2 (z)            \nonumber  \\
&& + (2z +k)(2 +3kz) C_0^{-3/2} Y(z) \biggr ] .
\end{eqnarray}

By now we have virtually completed our task of reducing $J_{23}$
in terms of one quadrature. By substituting Eqs. (\ref{Def-Phi4}),
(\ref{Fin-Phi2}), (\ref{Phi1}), and (\ref{Fin-Phi3}) into Eq. (\ref{J23}), 
we reorganize our result for $J_{23}$ in the following form:
\begin{eqnarray} \label{equation4}
J_{23} = 8 [P_0 + (k^2 -1)P_1 +2P_2 -kP_3 ] ,
\end{eqnarray}
where
\begin{eqnarray}  \label{P0}
P_0 =  &~& 8 k^2/3 \ln |k/2| -4/3           \nonumber \\
&+& [2 \ln 2 -1 + \ln |ab|] ab \ln ^2 |b/a|   \nonumber \\
&+&2 [k \ln2 - b (4k^2 - 2k +1)/3k ] \ln b    \nonumber \\ 
&-&2 [k \ln2 - a (4k^2 + 2k +1)/3k ] \ln |a| ,    
\end{eqnarray}
\begin{eqnarray}   \label{P1}
P_1 = \int_{-a}^b dz \frac{1}{C_0} \ln |z| ,
\end{eqnarray}
\begin{equation}   \label{P2}
P_2 = \int_{-a}^b dz \frac{1}{z}  [\lambda W_1 (z)
-(b+z)(z-a) W_2 (z) ] \ln |z| ,
\end{equation}
and
\begin{eqnarray} \label{Def-P3}
P_3 = \int_{-a}^b dz (k+ 2z)(2 +3kz) C_0^{-3/2} Y(z) \ln |z|.
\end{eqnarray}
It turns out that the integral in the expression for $P_3$ can be 
further refined. Indeed, it is shown in Appendix B that
\begin{eqnarray} \label{P3}
P_3 = &~& \frac{10}{3} (2k \ln|k/2| -1/k)     
+4 (b \ln ^2 b - a \ln ^2 |a| )     \nonumber \\
&+&\frac{4}{3} (3 \ln |2/k| -5 +1/k^2 )
(b \ln b - a \ln |a| )      \nonumber  \\
&+&\frac{1}{3k}(b \ln b + a \ln |a|) 
+\frac{1}{k}(k^2 -1) P_1 .
\end{eqnarray}
Remarkably, it can be seen from Eq. (\ref{P3}) that
the combination form of $(k^2 -1)P_1-kP_3$, 
the form in which $P_1$ and $P_3$ enter the right hand side 
of Eq. (\ref{equation4}), can be fully
reduced to an explicit form with no more integral in it.
In this light we substitute Eqs. (\ref{P0}) and (\ref{P3}) into 
Eq. (\ref{equation4}) to obtain the final expression for $J_{23}$
as the following: 
\begin{eqnarray} \label{Fin-J23}
J_{23} = 8 [ g (k) + 2 P_2 ],
\end{eqnarray}
where
\begin{widetext}
\begin{eqnarray}
g(k)=&& 2 +4k^2 \ln |2/k|+ab \ln|ab| \ln^2 |b/a|  
 +ab (2\ln 2-1) \ln ^2|b/a|   
+4k(a \ln^2|a| - b \ln^2 b)   \nonumber \\
&& +[2k^2 +9k/2 -2/k -2k (k+1) \ln 2 +4kb \ln k ] \ln b   \nonumber \\ 
&& +[2k^2 -9k/2 +2/k +2k (1-k) \ln 2 -4ka \ln k ] \ln |a| . 
\end{eqnarray}
\end{widetext}

\section{I(k) expressed in terms of one quadrature; and
further in terms of series}

Further substitution of Eq. (\ref{Fin-J1}) and Eq. (\ref{Fin-J23})
into Eq. (\ref{I-123}) then leads to
\begin{eqnarray} \label{Fin-I(k)}
I(k) = -\frac{1}{2k^2} [g_0 (k) + 2 P_2 ],
\end{eqnarray}
where
\begin{eqnarray} 
g_0 (k) 
=&~& 2k (a \ln ^2|a| -b \ln ^2 b) +ab \ln |ab| \ln ^2 |b/a| \nonumber \\
&-&2 (b \ln b -a \ln |a| -k -k \ln |k/2|)^2   \nonumber \\
&+& 2k^2 (1 + \ln ^2|k/2|) .
\end{eqnarray}
As a consequence, we have accomplished reducing $I(k)$ in
terms of one quadrature. 
Equation (\ref{Fin-I(k)}) will be further refined
in the next section into the following final form:
\begin{eqnarray}  \label{reducedI(k)}
I (k) &=& \frac{1}{2 k^2}  \biggl [
b \ln^2  \biggl | \frac{b}{a} \biggr | \biggl (\frac{1}{3}
a \ln \biggl |\frac{b}{a} \biggr | + b \biggr )   \nonumber \\
&+& \int_{a^2}^{b^2}
dx \frac{1}{x - a^2} \ln \biggl |\frac{x}{a^2} \biggr |
\biggl (\frac{1}{2} ab \ln \biggl |\frac{x}{b^2} \biggr |
-k \biggr ) \biggr ] ,
\end{eqnarray}
or, 
\begin{eqnarray}  \label{dimensionless}
I (k) &=& \frac{1}{2 k^2}   \biggl [
a \ln^2  \biggl | \frac{a}{b} \biggr | \biggl (\frac{1}{3}
b \ln \biggl |\frac{a}{b} \biggr | + a \biggr )     \nonumber \\
&+& \int_{1}^{a^2/b^2}
dx \frac{1}{1 - x}
\biggl (\frac{1}{2}ab \ln \biggl |\frac{a^2}{b^2x} \biggr | -k \biggr )
\ln x \biggr ] .      \nonumber \\
\end{eqnarray}
The resulting expression for $I(k)$
is amazingly simple, in constract
to its original six-fold integral form.
Equation (\ref{intro}) in the Introduction is obtained from
Eq. (\ref{reducedI(k)}) together with Eq. (\ref{Def-I)}).

The one quadrature in Eq. (\ref{reducedI(k)}) presents no
numerical difficulty in applications. This is particularly so with the
aid of the expansion forms in the limiting cases obtained in 
Sec. VII below. But it turns out that we have even a better choice.
In this section, a series representation for $I(k)$ is
developed. To this end, we rewrite Eq. (\ref{dimensionless}) as
\begin{widetext}
\begin{equation}  \label{dimensionless1}
I (k)= \frac{1}{2 k^2}  \biggl [
a \ln^2  \biggl | \frac{a}{b} \biggr | \biggl (\frac{1}{3}
b \ln \biggl |\frac{a}{b} \biggr | + a \biggr )
+\biggl ( ab\ln \biggl | \frac{a}{b} \biggr | -k \biggr )
\frac{\pi^2}{6} +ab \zeta (3)
+\int_{0}^{a^2/b^2}
dx \frac{1}{1 - x}
\biggl (\frac{1}{2}ab \ln \biggl |\frac{a^2}{b^2x} \biggr | -k \biggr )
\ln x \biggr ]  .
\end{equation}
\end{widetext}
In obtaining Eq. (\ref{dimensionless1}) we have employed identities
\begin{eqnarray}   \label{id1}
\int_0^1 dx \frac{1}{1-x} \ln x = - \frac{\pi^2}{6} ,
\end{eqnarray}
and
\begin{eqnarray}
\int_0^1 dx \frac{1}{1-x} \ln^2 x = 2 \zeta (3) .
\end{eqnarray}
We now employ series representation $1/(1-x)=\sum_{l=0}^\infty x^l$ in
Eq. (\ref{dimensionless1})
to obtain
\begin{equation}  \label{series}
I (k)= \frac{1}{2 k^2} \biggl [f_0(k) + \sum_{n=1}^{\infty}
f_n(k) \biggr ] ,
\end{equation}
where
\begin{eqnarray}
f_0 (k) &=&
a \ln^2  \biggl | \frac{a}{b} \biggr | \biggl (\frac{1}{3}
b \ln \biggl |\frac{a}{b} \biggr | + a \biggr )
-\frac{\pi^2}{6}k +ab \zeta (3) - \frac{a^3}{2b}   \nonumber \\
&+&\biggl ( \frac{\pi^2}{6}ab
+2k \ln \biggl |\frac{2k}{b^2} \biggr | \biggr )
\ln \biggl | \frac{a}{b} \biggr |    \nonumber  \\
&+& \biggl (\frac{a}{b} \biggr )^2 \biggl [1+
\biggl ( \frac{a}{2b} \biggr )^2 \biggr ]
\biggl [ k - ab \biggl ( \frac{1}{2} + \ln \biggl |\frac{b}{a}
\biggr | \biggr ) \biggr ] ,   \nonumber \\
\end{eqnarray}
and
\begin{equation}  \label{f1}
f_n(k) =
\biggl ( ab \ln \biggl |\frac{a}{b} \biggr |
+k - \frac{ab}{n+2} \biggr ) \frac{1}{(n+2)^2} \left ( \frac{a}{b}
\right )^{2(n+2)}.
\end{equation}
We emphasize that the series on the right hand side of
Eq. (\ref{series}) converge for all $k \geq 0$ and in effect the value of
$I(k=0)=-1$ can be equally (and rather easily indeed) obtained from
Eq. (\ref{series}) by employing the following identity,
\begin{equation}
\sum_{n=1}^\infty \frac{1}{n^p} = \zeta (p),
\end{equation}
with $\zeta (2)=\pi^2/6$ explicitly.
The series representation of Eq. (\ref{series}) for $I(k)$ 
should be particularly useful for analytical purposes.

\section{Derivation of Eq. (\ref{reducedI(k)}) and proof
of the equivalence of it and the corresponding one of Eq. (29) 
in Ref. \cite{Engel}}

It is by no means evident that $I(k)$ given as Eq. (\ref{Fin-I(k)})
is equal to the corresponding one of Eq. (29) in Ref. \cite{Engel}. 
In fact the present author had certain reservations about 
Eq. (29) of Ref. \cite{Engel} for the mathematical assumption
used in deriving it, and was convinced of its correctness
only after a numerical check with Eq. (\ref{Fin-I(k)}). 
The numerical results from both
of the expressions (before the rigorous proof of their equivalence
was found) are shown in Fig. 1. They attain the same illustration. 
Besides, it can be derived analytically from Eq. (\ref{Fin-I(k)}) that
$I(k=2)=-\pi^2/24$ [see also Eq. (\ref{k=2result1}) and
Eq. (\ref{k=2result2}) below], which is the same as that 
derived from Eq. (29) of Ref. \cite{Engel}.
In this section we prove their equivalence by refining both of 
them into the form of Eq. (\ref{reducedI(k)}).
\begin{figure}
\unitlength1cm
\begin{picture}(5.0,7.0)
\put(-5.2,-4.0){\makebox(7.0,8.0){
\includegraphics{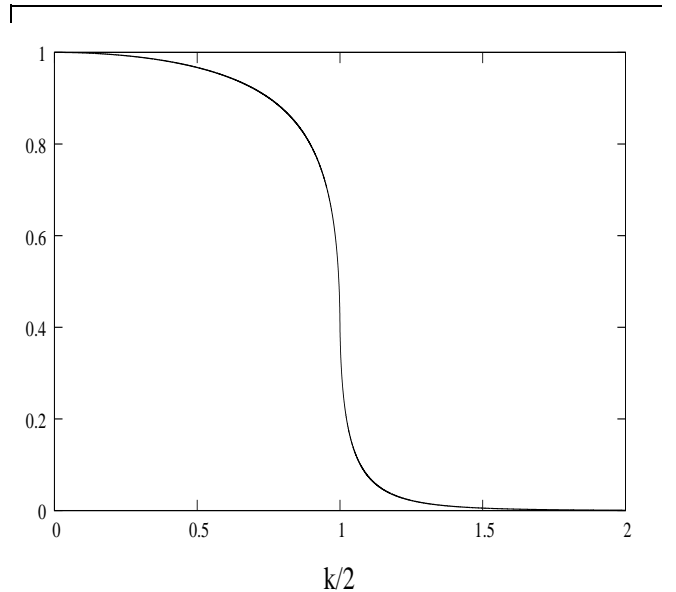}
}}
\end{picture}
\caption{The result for the quantity $- I(k)$,
calculated with both of Eq. (\ref{Fin-I(k)})
of this paper and Eq. (29) of Ref. \cite{Engel}.}
\label{figur1}
\end{figure}

We first refine the quantity given in
Eq. (29) of Ref. \cite{Engel} into the form of Eq. (\ref{reducedI(k)}). 
For the purpose of convenience, we call it $I^{EV}(Q)$ (with $Q=k/2$).
To this end, we make the use of the following identity,
\begin{eqnarray}
\int_0^Q dx \frac{1-x^2}{x^2}  && \ln ^3 \biggl | \frac{1+x}{1-x} \biggr | 
= -(Q+ \frac{1}{Q} )\ln ^3 \biggl | \frac{1+Q}{1-Q} \biggr |   \nonumber \\ 
 && + 6 \int_0^Q dx \frac{1+x^2}{x(1-x^2)}\ln ^2 
\biggl | \frac{1+x}{1-x} \biggr |  ,
\end{eqnarray}
to rewrite Eq. (29) of Ref. \cite{Engel} as
\begin{eqnarray}   \label{engel}
I^{EV} (Q) = \omega_1      
&+& \omega_2 \int_0^Q dx \frac{1+x^2}{x(1-x^2)} 
\ln ^2 \biggl | \frac{1+x}{1-x} \biggr |   \nonumber \\  
&+&\omega_3 
\int_0^Q dx \frac{1-x^2}{x^2}     
\ln ^2 \biggl | \frac{1+x}{1-x} \biggr | ,
\end{eqnarray} 
where
\begin{eqnarray}
\omega_1 = - \frac{1- Q^4}{16 Q^3} \ln ^3 \biggl |
\frac{1+Q}{1-Q} \biggr | , ~~
\omega_2 =
\frac{1-Q^2}{4Q^2}  ,
\end{eqnarray}
\begin{eqnarray}
\omega_3 =
-\frac{1}{8} \biggl (\frac{1}{Q} +\frac{1 -Q^2}{2 Q^2 } \ln  \biggl |
\frac{1+Q}{1-Q} \biggr | \biggr ) .
\end{eqnarray}
We next make the variable transform
$x=(y-a)/(y+a)$  
in Eq. (\ref{engel}) to obtain 
\begin{eqnarray}  \label{vosko}
I^{EV} (Q) = \omega_1 + \int_{a}^b dy &[&\omega_2 (y^4 -a^4) 
+ 8 \omega_3 a^2 y^2 ~]        \nonumber \\
&~& \frac{1}{y (y^2 - a^2)^2} \ln ^2 \biggl |\frac{y}{a} \biggr | .
\end{eqnarray}
Further algebra then leads to
\begin{eqnarray}   
I^{EV} (Q) &=& \omega_1 - \biggl [\frac{\omega_2}{3} 
\ln \biggl |\frac{b}{a} \biggr | + \frac{2a^2\omega_3}{k} \biggr ]
\ln ^2 \biggl |\frac{b}{a} \biggr |    \nonumber \\
&+& \int_{a^2}^{b^2} dy \biggl [ 
\frac{\omega_2}{4} \ln \biggl |\frac{y}{a^2} \biggr |
+ \frac{2 \omega_3 a^2}{y} \biggr ]        
\frac{1}{ y - a^2} \ln \biggl |\frac{y}{a^2} \biggr | . \nonumber \\
\end{eqnarray}
The remaining derivation to obtain the form of Eq. (\ref{reducedI(k)})
is routine.

We next refine $I(k)$ of Eq. (\ref{Fin-I(k)}) 
into Eq. (\ref{reducedI(k)}). To this end, we carry out
partial integration on the right hand side of Eq. (\ref{P2})
to rewrite it as
\begin{eqnarray} \label{Fin-Q}
P_2 = k [b \ln ^2 b \ln |2b/k| -a \ln ^2 |a| \ln |2a/k| ] +{\bar P}_2,
\end{eqnarray}
with
\begin{equation}
{\bar P}_2 = \frac{1}{2} \int_{-a}^b dz  [(2z-k) W_1 (z)   
+(2z +k) W_2 (z) ] \ln ^2 |z| .   
\end{equation}
With some straightforward algebra, ${\bar P}_2$ can be put into 
the form
\begin{eqnarray}
{\bar P}_2 = (t_1 -4kt_2 -4k t_3 -t_4 )/8 ,
\end{eqnarray}
where
\begin{eqnarray} 
t_1 = \int _{a^2} ^{b^2} dy \ln ^2 y \ln |y-a^2 |,
\end{eqnarray}
\begin{eqnarray}
t_2 =\int _{|a|} ^{b}dy \ln ^2 y \ln \biggl | \frac{y+a}{y-a} \biggr |,
\end{eqnarray}
\begin{eqnarray}
t_3 =\int _{|a|} ^{b}dy \ln ^2 y \ln \biggl |\frac{y+b}{y-b} \biggr |,
\end{eqnarray}
and
\begin{eqnarray}
t_4 = \int _{a^2} ^{b^2} dy \ln ^2 y \ln |y-b^2 | .
\end{eqnarray}
The next manipulation is the key to the procedure.
For $t_1$ we perform further partial integration to obtain
\begin{eqnarray}
t_1 &=&  [b^2(d_4^2+1) -a^2(d_1^2 +1)] \ln|2k|  \nonumber \\
&-&\int_{a^2}^{b^2} dy \frac{1}{y-a^2}
[ y (\ln^2 y - 2 \ln y +2) -a^2 (d_1^2 +1)] , \nonumber \\
\end{eqnarray}
where 
\begin{eqnarray}
d_1 &=&  2\ln |a| -1 , ~~~d_2 = 2(\ln |a| -1),   \nonumber \\
d_3 &=& 2(1 -\ln b) , ~~~d_4= - 2\ln b +1  , 
\end{eqnarray}
with $d_2$, $d_3$ for later quotations.
Further algebra then yields
\begin{equation}   \label {t1}
t_1 =  b^2 (d_4^2 +1 ) \ln |2k|    
- (d_1^2 +1 )(2k + a^2 \ln |2k|) -\Xi_1 ,
\end{equation}
where
\begin{equation}
\Xi_n =\int_{a^2}^{b^2} dy \frac{y^{2-n}}{y -a^2} \biggl ( 2 d_n
+ \ln \biggl |\frac{y}{a^2} \biggr | \biggr )
\ln \biggl |\frac{y}{a^2} \biggr | ,
\end{equation}
with $\Xi_2$, $\Xi_3$, and $\Xi_4$ for later quotations.
Similarly we can obtain
\begin{equation}    \label{t2}
t_2 = \frac{1}{4} [b(d_3^2+4)  \ln|2/k|         
- a (d_2^2 +4 ) \ln |2a^2/k|   
+a \Xi_2 ].
\end{equation}
For $t_4$ we first perform partial integration to get
\begin{eqnarray}  
t_4 &=&  [b^2(d_4^2+1) -a^2(d_1^2 +1)] \ln|2k|  \nonumber \\
&-&\int_{a^2}^{b^2} dx \frac{1}{x-b^2}
[ x (\ln^2 x - 2 \ln x +2) -b^2 (d_4^2 +1)] .\nonumber \\
\end{eqnarray}
We then make further variable tranform $x=a^2b^2/y$ to
get
\begin{equation}  \label{t4}
t_4 =  (d_4^2 +1)( b^2 \ln |2k| -2k)         
-a^2 (d_1^2  +1 ) \ln |2k|   
+a^4b^2 \Xi_4 .
\end{equation}
Similarly we can obtain
\begin{equation}   \label{t3}
t_3 =  \frac{1}{4} [a (d_2^2 +4 ) \ln |k/2| +   
b (d_3^2 +4) \ln |2b^2/k| - a^2b  \Xi_3] . 
\end{equation}
Substitution of Eqs. (\ref{t1}), (\ref{t2}), (\ref{t4}), 
(\ref{t3}) into Eq. (\ref{Fin-Q}) yields
\begin{eqnarray}  \label{P2T}
P_2 = &~& k (\ln b -1)( \ln b +2b \ln |2b/k|)    \nonumber \\ 
&-& k (\ln |a| -1)( \ln |a|+2a \ln |2a/k|)
- T/2,   \nonumber \\
\end{eqnarray}
where
\begin{eqnarray}  
T= [\Xi_1 + ka\Xi_2 -kba^2 \Xi_3 +a^4b^2 \Xi_4]/4  , 
\end{eqnarray}
which has the following explicit expression:
\begin{widetext}
\begin{eqnarray}   \label{T}
T= [2k \ln |ab| + b(b-2a +2a \ln b) \ln |b/a |
-2k  -2ab/3 \ln^2 |b/a |] \ln |b/a |   
+  \int_{a^2}^{b^2}
dx \frac{1}{x - a^2} \ln \biggl |\frac{x}{a^2} \biggr |
\biggl (\frac{1}{2} ab \ln \biggl |\frac{x}{b^2} \biggr |
-k \biggr ) . \nonumber \\
\end{eqnarray}
One then substitutes Eq. (\ref{P2T}) together with Eq. (\ref{T}) 
into Eq. (\ref{Fin-I(k)})
for $I(k)$. 
The remaining derivation to get the form of Eq. (\ref{reducedI(k)})
becomes routine.

\section{Expansions in the limiting cases}

The limiting structures of $\Pi_1 (k, 0)$ at $k=0$, $k=2$, and
at large $k$ have various physical implications
and deserve particular attention. 
In virtue of our proof of the correctness of the 
main conclusion of Engel and Vosko's theory
in Ref. \cite{Engel}, there is no doubt that the expansions
obtained by them in the limiting cases [i.e., Eqs. (30), 
(31), and (32) in Ref. \cite{Engel}] must also be right.
Indeed they are all confirmed with calculations based on
Eq. (\ref{dimensionless}).
We sketch the calculations below. 

It turns out that the expression (\ref{dimensionless}) is rather suitable 
for deriving the expansions. 
We first consider the cases of $k \to 0$ and $k \to \infty$, 
To this end, we make the variable 
transform
$x= [(1-y)/(1+y)]^2$
in Eq. (\ref{dimensionless}) and bring it into another form,
\begin{equation}  \label{expansion}
I(k) =\frac{1}{2k^2} \biggl [ 
b \ln ^2 \biggl |\frac{b}{a} \biggr | \biggl ( \frac{1}{3} a 
\ln \biggl |\frac{b}{a} \biggr | + b \biggr )  
+ 2 \int_0^{Q} dy
\biggl ( \frac{1}{y} + \frac{2}{1-y} \biggr ) \ln \biggl |
\frac{1+y}{1-y} \biggr |   
\biggl (  ab \ln \biggl |\frac{a(1+y)}{b(1-y)} \biggr | 
-k \biggr )  \biggr ] .
\end{equation}
We act the operator of
$\frac{1}{2} (1 - Q^2) \frac{\partial}{Q \partial Q} +1$
on the quantity $I(k) Q^2$. With some algebra,
one can obtain
\begin{equation}   
\biggl [\frac{1}{2} (1 - Q^2) \frac{\partial}{Q \partial Q} 
+1 \biggr ] [I (k)Q^2]
=\frac{1}{4Q} \biggl [b \ln ^2 \biggl |\frac{b}{a} \biggr | 
-2 \int_0^Q dy \biggl ( \frac{1}{y}+\frac{2}{1-y} \biggr ) \ln \biggl |
\frac{1+y}{1-y} \biggr | \biggr ] .
\end{equation}
\end{widetext}
One then acts further the operator of $(1/Q)$$\partial / \partial Q$
(which is of course just $2 \partial / \partial Q^2$) on the both 
sides of the preceding expression. This action yields the following
result,
\begin{equation}   \label{opera}
\frac{\partial}{Q\partial Q}\biggl [\frac{1}{2} 
(1 - Q^2) \frac{\partial}{Q \partial Q}
+1 \biggr ] [I (k)Q^2]
=\frac{1}{4Q^3} F(Q) ,
\end{equation}
where
\begin{eqnarray}
F(Q) =&-& \ln \biggl |\frac{b}{a} \biggr |
\biggl ( \ln \biggl |\frac{b}{a} \biggr | + 2 \biggr )
\nonumber \\
&+&2 \int_0^Q dy \biggl (
\frac{1}{y} + \frac{2}{1-y} \biggr ) \ln \biggl |
\frac{1+y}{1-y} \biggr | .
\end{eqnarray}
It can be readily shown that
\begin{equation}  \label{F(Q)1}
F(Q \to 0) = 4\sum_{n=0}^\infty \frac{1}{2n+3} \biggl [ 2 \Psi (n)
+\frac{1}{2n+3} -1 \biggr ] Q^{2n+3} ,
\end{equation}
and
\begin{equation}  \label{F(Q)2}
F(Q \to \infty) = 4\sum_{n=0}^\infty \frac{1}{2n+3} \biggl 
[ 2 \Psi (n) +\frac{1}{2n+3} -1 \biggr ] \frac{1}{Q^{2n+3}} ,
\end{equation}
where
\begin{eqnarray}   \label{Euler}
\Psi (n) = \sum_{l=0}^n \frac{1}{2l +1} .
\end{eqnarray}
In passing we note the remarkable symmetry in the above two forms.
For $k \to 0$ we have, from Eq. (\ref{opera})
and Eq. (\ref{F(Q)1}),
\begin{eqnarray}   \label{opera2}
&& \frac{\partial}{Q\partial Q}\biggl [\frac{1}{2} 
(1 - Q^2) \frac{\partial}{Q \partial Q}
+1 \biggr ] [I (k)Q^2]   \nonumber \\
&&=\sum_{n=0}^\infty \frac{1}{2n+3} \biggl [ 2 \Psi (n) 
+\frac{1}{2n+3} -1 \biggr ] Q^{2n} .
\end{eqnarray}
With Eq. (\ref{opera2}), it is straightforward to determine
the expansion for $I(k \to 0)$. 
To this end, we write
\begin{eqnarray}   \label{I-form}
I(k \to 0) = -1 + \sum_{n=1}^\infty c_n Q^{2n} ,
\end{eqnarray}
where $c_n$ are the coefficients we have to determine. To write
the preceding form, we have made the use of the truth of $I(k=0)=-1$ 
which can be readily obtained from Eq. (\ref{expansion}) and  
is in fact well known \cite{DuBois,Kleinman,Chevary,Engel}. 
The fact that the expansion retains only even powers of $Q$ is 
evident from Eq. (\ref{opera2}).
[In fact $I(k)$, which is defined physically only for
$k \ge 0$, is an even function if extended to the range
of $k <0$. This fact can be best appreciated from Eq. (\ref{engel}),
but also in another interesting way. Upon
making the variable tranform $x=1/y$ in Eq. (\ref{dimensionless}),
one can bring it into yet another form,
\begin{eqnarray}  
I (k)= &~& \frac{1}{2 k^2}  \biggl [
b \ln^2  \biggl | \frac{b}{a} \biggr | \biggl (\frac{1}{3}
a \ln \biggl |\frac{b}{a} \biggr | + b \biggr )    \nonumber \\
&+&\int_{1}^{b^2/a^2}
dx \frac{1}{x - 1}
\biggl (\frac{1}{2}ab \ln \biggl |\frac{a^2x}{b^2} \biggr | -k \biggr )
\ln x \biggr ] .       \nonumber \\
\end{eqnarray}
But, on the other hand, the right hand side of the above equation 
can be exactly obtained
from that of Eq. (\ref{dimensionless}) by merely changing $k$ to $-k$.
Thus $I(k)= I(-k)$.]
We then substitute Eq. (\ref{I-form}) 
into the left hand side of Eq. (\ref{opera2}). By equating the
coefficients of the same powers of $Q^2$ on both sides of the 
resulting equation, we obtain, for $n \ge 1$,
\begin{eqnarray}  
&&(n+1)(n+2) c_{n+1} - n(n+1) c_n    \nonumber \\
&& =\frac{1}{2(2n+3)}
\biggl [ 2 \Psi (n) + \frac{1}{2n+3} -1 \biggr ] ,
\end{eqnarray}
or further
\begin{equation}
n(n+1) c_{n}  = \frac{1}{2}\sum_{l=1}^{n} \frac{1}{2l+1}
\biggl [ 2 \Psi (l-1) + \frac{1}{2l+1} -1 \biggr ] .
\end{equation}
By the use of Eq. (\ref{Euler}), we have
\begin{equation}
n(n+1) c_{n}= \frac{1}{2}\sum_{l=1}^{n} [\Psi (l) - \Psi (l-1) ] 
[\Psi (l) + \Psi (l-1) -1  ] .
\end{equation}
After carrying out the above summation we have, with the aid
of the fact $\Psi(0)=1$,
\begin{eqnarray}
c_{n}  = \frac{\Psi (n) [ \Psi (n) -1 ]}{2n(n+1)} ,
\end{eqnarray}
and accordingly
\begin{eqnarray} \label{k=0final}
I (k \to 0) = -1 + \frac{1}{2}\sum_{n=1}^\infty \frac{\Psi (n) 
[ \Psi (n) -1 ]} {n(n+1)} Q^{2n} .
\end{eqnarray}
Equation (\ref{k=0final}) is the same as Eq. (30) of Ref. \cite{Engel}. 

In particular, the leading two terms in Eq. (\ref{k=0final}) which are
\begin{eqnarray} \label{leadingterm}
I (k \to 0) = -1 + \frac{1}{9} Q^2 + ... ,
\end{eqnarray}
had been obtained numerically by Kleinman and Lee \cite{Kleinman} and
later by Chevary and Vosko \cite{Chevary} (who also obtained the next
order $46 Q^4/675$ correctly). 
The form in 
Eq. (\ref{leadingterm}) yields the following leading gradient 
correction \cite{Kohn,Ma,Sham,Langreth1,Langreth,Dreizler} 
to the local density approximation of Kohn-Sham 
exchange energy density \cite{Kleinman,Chevary,Engel}
\begin{eqnarray} \label{gradient}
\bigtriangleup E_x^{(1)} = - \frac{5e^2}{216 \pi (3 \pi^2)^{1/3}} 
\int d {\bf r}
\frac{[\bigtriangledown n({\bf r})]^2}{[n({\bf r})]^{4/3}} .
\end{eqnarray}
The leading gradient correction to the Kohn-Sham exchange energy 
was investigated originally by Sham \cite{Sham1},  
but he got the coefficient to be $-7e^2/[432 \pi (3 \pi^2)^{1/3}]$ 
instead by employing screened Coulomb potential with screening going 
to zero (see also Ref. \cite{Kleinman2}). Later Gross and 
Dreizler \cite{Gross} (see also Ref. \cite{Mohammed})
got the same result as Sham's, essentially employing also screened 
Coulomb potential (with screening going to zero). This well-known 
controversy has been one of the major causes inspiring the 
investigations of the exact structures of $\Pi_1 ({\bf k} ,0)$, and has
been effectively elucidated in Refs. \cite{Kleinman,Chevary,Engel}. Our
result based on the rigorous derivation helps to {\em give a final 
settlement of the controversy itself}. Questions such as 
whether $\bigtriangleup E_x^{(1)}$ in Eq. (\ref{gradient}) 
or the one by Sham should be added to the 
type of Ma-Brueckner's gradient correction to the correlation energy 
(to get the Kohn-Sham exchange-correlation energy beyond the
local density approximation) remain open
\cite{Kleinman,Langreth}.

For $k \to \infty$, we have, from Eq. (\ref{opera})
and Eq. (\ref{F(Q)2}),
\begin{eqnarray}   \label{opera3}
&& \frac{\partial}{Q\partial Q}\biggl [\frac{1}{2}
(1 - Q^2) \frac{\partial}{Q \partial Q}
+1 \biggr ] [I (k)Q^2]   \nonumber \\
&&=\sum_{n=0}^\infty \frac{1}{2n+3} \biggl [ 2 \Psi (n)
+\frac{1}{2n+3} -1 \biggr ] \frac{1}{Q^{2n+6}} .
\end{eqnarray}
It is not difficult to see from Eq. (\ref{opera3}) that the expansion
of $I(k)$ for large $k$ commences with $O(1/Q^6)$, and accordingly
it assumes the following general form,
\begin{equation}
I(k \to \infty) = \sum_{n=3}^\infty {\tilde c}_n \frac{1}{Q^{2n}} .
\end{equation}
Substituting this form into Eq. (\ref{opera3}), we obtain
\begin{eqnarray} 
&&(n+2)(n+3) {\tilde c}_{n+3} - (n+1)(n+2) {\tilde c}_{n+2}    \nonumber \\
&& =\frac{1}{2(2n+3)}
\biggl [ 1 - \frac{1}{2n+3} -2 \Psi (n) \biggr ] ,
\end{eqnarray}
for $n \ge 1$. Similar procedure to that used above for 
the case of $k \to 0$ can be employed to obtain
\begin{equation}
{\tilde c}_n = - \frac{\Psi (n-2) [ \Psi (n-2) - 1]}
{2 (n-1) n} .
\end{equation}
Thus 
\begin{equation}   \label{large-k}
I( k \to \infty )= - \frac{1}{2}\sum_{n=1}^\infty
\frac{\Psi (n) [ \Psi (n) - 1]}
{(n+1) (n+2)} \frac{1}{Q^{2(n+2)}} ,
\end{equation}
confirming Eq. (31) in Ref. \cite{Engel}, (there is a sign error
in the first identity of that equation.) We note that the leading
two terms of Eq. (\ref{large-k}) had been reported by Geldart
and Taylor \cite{Geldart1} (see also Ref. \cite{Rasolt}); 
the leading term had
also been obtained by Kleinman \cite{Kleinman1,Kleinman2}.

We next give the expansion near $k=2$. 
With the aid of Identity (\ref{id1}),
the characteristic property
of $I(k=2)= -\pi^2/24$ and the leading singular term (meaning the most
divergent term upon derivative with respect to $k$) can be 
immediately seen from Eq. (\ref{dimensionless}) to be:
\begin{eqnarray}    \label{k=2result1}
I (k \to 2) = -\frac{\pi^2}{24} + \frac{1}{12} a \ln ^3 \biggl 
|\frac{a}{2} \biggr | .
\end{eqnarray}  
The conclusion drawn in Ref. \cite{Engel} that the singularity
of $\Pi_1 ({\bf k}, 0)$ dominates that of $\Pi_0({\bf k}, 0)$
is corroborated.
In general, we obtain
\begin{widetext}
\begin{eqnarray}   \label{k=2result2}
I(k \to 2) = \frac{1}{8} \biggl \{
&-&\frac{\pi^2}{3} + \biggl (2 \zeta (3) - \frac{\pi^2}{3} 
\biggr )a +\biggl (3 \zeta (3) - \frac{\pi^2}{6} 
+\frac{1}{2} \biggr ) a^2 + \biggl (4 \zeta (3) 
- \frac{\pi^2}{24} \biggr ) a^3 +\biggl (5 \zeta (3) 
+ \frac{11\pi^2}{144}           
-\frac{37}{96} \biggr ) a^4     \nonumber \\
&+& \biggl (6 \zeta (3) 
+ \frac{37\pi^2}{192}
-\frac{125}{192} \biggr ) a^5  + ...\nonumber \\
&+& \frac{1}{3} \sum_{n=1}^\infty
(n+1) a^n \ln^3 \biggl |\frac{a}{2} \biggr |   
+ a^2  \sum_{n=0}^\infty
e_n a^n  
\ln^2 \biggl |\frac{a}{2} \biggr |      
+ a \biggl [ \frac{\pi^2}{3} + \biggl (\frac{\pi^2}{2} 
-1 \biggr ) a + \sum_{n=2}^\infty
r_{n+1} a^n \biggr ] 
\ln \biggl |\frac{a}{2} \biggr |   
 \biggr \} ,
\end{eqnarray}
where
\begin{equation}
e_n= 2(n+1) + \sum_{l=0}^n \frac{l-n}{(l+1)(l+2)}\frac{1}
{2^{l+1}} ,
\end{equation}
and
\begin{equation}
r_n= \frac{\pi^2}{6} (n+1) + 7 (n-1) - 8 \phi (n-2) + \sum_{m=3}^n 
\frac{(n-m+1)}{(m-1)(m-2)} 2^{3-m} \biggl [ -1 + \phi (m-1) +
\frac{3(m-1)^2 -1 }{m (m-1)(m-2)} - \sum_{l=1}^m \frac{1}{l} 2^l 
\biggr ] ,
\end{equation}
\end{widetext}
with
$\phi (m) = \sum_{l=0}^m 1/(l+1)$ .
In particular, $e_0=2$, $e_1=15/4$, $e_2=131/24$, and $e_3=229/32$;
$r_3=2\pi^2/3$, $r_4=$$5 \pi^2/6$$+13/8$, and $r_5=$$\pi^2$$+173/48$.
Equation (\ref{k=2result2}) confirms Eq. (32) in Ref. \cite{Engel} 
in virtue of the following identities,
\begin{eqnarray}
&&\int_0^1 dx \frac{1-x^2}{x^2} \ln^3 \biggl | 
\frac{1+x}{1-x}\biggr |    \nonumber \\
&&= 6 \biggl [ \int_0^1 dy \frac{1}{1-y} \ln^2 y
- \int_0^1 dy \frac{1}{1+y} \ln^2 y \biggr ]   \nonumber \\
&&=3 \zeta (3) ,
\end{eqnarray}
where $\zeta (3)=1.202056903...$.
We would like to point out in this connection
that Eq. (21) in Ref. \cite{Glasser} is at variance
with Eq. (32) in Ref. \cite{Engel} and accordingly also with
Eq. (\ref{k=2result2}) above.

\appendix

\section{Verification for Eqs. (\ref{psi1/psi2}), (\ref{psi3/psi4}), 
and (\ref{phi1/phi2})}

It is worth pointing out first the following fact,
\begin{equation}
(1+k)z + b \ge 0, ~~and~~ (k-1)z +a \ge 0 ,
\end{equation}
which can be readily verified for $-a$ $\le z$ $\le b$.
Accordingly,
\begin{equation}  \label{R1R2}
\sqrt{R(z, b)} =(1+k)z + b ; ~~  \sqrt{R(z, -a)} = (k-1)z +a .
\end{equation}
A direct substitution of Eq. (\ref{ABC}) and Eq. (\ref{R1R2}) 
into Eq. (\ref{psi1}) and Eq. (\ref{psi2}), respectively, leads to
\begin{eqnarray}
\psi_1(z) = 4b^2 z(z+a) , ~~and~~
\psi_2(z) = k^2 z(z+a) ,
\end{eqnarray}
for $z > 0$; and
\begin{eqnarray}
\psi_1(z) = k^2 z (z-b) ,~~and~~
\psi_2(z) = 4a^2 z(z-b) ,
\end{eqnarray}
for $z < 0$. Equation (\ref{psi1/psi2}) is hence verified.
Equation (\ref{phi1/phi2}) can be verified in the same manner.

We next verify Eq. (\ref{psi3/psi4}). Once again we make  
a direct substitution of Eq. (\ref{ABC}) and Eq. (\ref{R1R2}) 
into Eq. (\ref{psi3}) and Eq. (\ref{psi4}), respectively.
By carrying out some algebra, we can obtain
\begin{eqnarray}
\psi_3(z)= \frac{1}{2k} &[& C_0^2 +2(k+1) C_0^{3/2} + k (k+4) C_0
\nonumber   \\
&+& 2(k^2 +k - 1) \sqrt{C_0} +k^2 - 1]  ,
\end{eqnarray}
and
\begin{eqnarray}
\psi_4(z)= \frac{1}{2k} &[& C_0^2 +2(k-1) C_0^{3/2} + k (k-4) C_0  
\nonumber  \\
&+& 2(-k^2 +k + 1) \sqrt{C_0} +k^2 - 1]  .
\end{eqnarray}
Both of these forms can be factorized, and the result is
\begin{equation}
\psi_3(z)= \frac{1}{2k} ( \sqrt{C_0} +1)^2 ( \sqrt{C_0} +k +1)
( \sqrt{C_0} +k-1) ,
\end{equation}
and
\begin{equation}
\psi_4(z)= \frac{1}{2k} ( \sqrt{C_0} -1)^2 ( \sqrt{C_0} +k +1)
( \sqrt{C_0} +k-1) .
\end{equation}
Equation (\ref{psi3/psi4}) is as a consequence established.

\section{Evaluation for $\eta_n$ in Eq. (\ref{eta-n})
and derivation for Eq. (\ref{P3})}

An efficient way to evaluate $\eta_n$ and $P_3$ is to introduce
the variable transform,
$z=\frac{1}{2k} (y^2 -1)$,
which casts Eq. (\ref{eta-n}) in the form
\begin{eqnarray}
\eta_n =\int_{|1-k| -1}^k
dy (y+1)^{n+1} \ln \biggl | \frac{y+2}{y} \biggr | .
\end{eqnarray}
The remaining calculation for $\eta_n$ is then routine.
The explicit expressions for $\eta_1$, $\eta_{-1}$, and $\eta_{-3}$ are
given as follows,
\begin{widetext}
\begin{equation} \label{I1/2}
\eta_1 = \frac{2}{3}[2k
+ (k^2 + k+1)b \ln |2b|
- (k^2 - k+1)a \ln |2a| - k (k^2 +3 )\ln k ] ,
\end{equation}
\begin{eqnarray} \label{I-1/2}
\eta_{-1} = 2 ( b \ln |2b| - a \ln |2a| - k \ln k ) ,
\end{eqnarray}
and
\begin{equation}  \label{I-3/2}
\eta_{-3} = 2 \biggl [ \frac{k}{k^2 -1} \ln k 
+\ln \biggl | \frac{k+1}{k-1} \biggr |
-\frac{b}{k+1} \ln |2b|  
+\frac{a}{1-k} \ln |2a| \biggr ].
\end{equation}

We next evaluate $P_3$ to verify Eq. (\ref{P3}). With the transform
$z=\frac{1}{2k} (y^2 -1)$, Eq. (\ref{Def-P3})
can be rewritten as
\begin{equation} \label{Fin-P3}
P_3 =\frac{1}{2k^2}\sum_{n=-1}^1 \gamma_n \int_{|1-k|}^{1+k}
dy y^{2n} \ln \biggl | \frac{y^2 -1}{2k} \biggr |
\ln \biggl | \frac{y+1}{y-1} \biggr | ,
\end{equation}
where
\begin{eqnarray}
\gamma_1=3,~~ \gamma_0 =3k^2 -2, ~~
\gamma_{-1} = k^2 -1 .
\end{eqnarray}
One then carries out partial integration to bring Eq. (\ref{Fin-P3}) 
into the following form,
\begin{equation}  \label{P3-App}
P_3 =\frac{1}{2k^2}\sum_{n=-1}^1  \gamma_n \frac{1}{2n+1}
\biggl [ (1+ k)^{2n+1} \ln b \ln \biggl | \frac{2b}{k} \biggr | 
-{\tilde k}^{2n+1} \ln |a| 
\ln  \biggl | \frac{{\tilde k} +1}{{\tilde k} -1} \biggr |
+ \Omega_n \biggr ] ,
\end{equation}
with ${\tilde k} =|1-k|$ and $\Omega_n$ defined as
\begin{eqnarray} 
\Omega_n =\int_{\tilde k}^{1+k} dy y^{2n+1} \biggl [  \frac{1}{y +1} 
\ln \biggl | \frac{2k}{(y+1)^2} \biggr |  
 - \frac{1}{y -1}
\ln \biggl | \frac{2k}{(y-1)^2} \biggr | \biggr ] ,
\end{eqnarray}
for $n=1, 0, -1$. The $\Omega_{-1}$ can be
evaluated as
\begin{equation}  \label{Omega-1}
\Omega_{-1}= h_0 (k)
-2k \int_{-a}^b dz \frac{1}{C_0} \ln |z| ,
\end{equation}
where
\begin{equation}
h_0 (k) =\ln |2k|  \ln  \biggl | \frac{a}{b} \biggr | 
+\ln^2 |2b| - \ln ^2 |{\tilde k} +1| +\ln ^2 k     
- \ln ^2 |{\tilde k} -1|.
\end{equation}
For $n=0, 1$, one has
\begin{eqnarray}
\Omega_n =\int_{{\tilde k}+1}^{2b} dx (x-1)^{2n+1}  \frac{1}{x}
\ln \biggl | \frac{2k}{x^2} \biggr |   
-\int_{{\tilde k}-1}^{k} dx (x+1)^{2n+1}  \frac{1}{x}
\ln \biggl | \frac{2k}{x^2} \biggr |  ,
\end{eqnarray}
or
\begin{equation}  \label{Omega-n}
\Omega_n =\sum_{m=1}^{2n+1} C_{2n+1}^m
\biggl [ (-)^m \int_{{\tilde k}+1}^{2b} dx x^{m-1}
\ln \biggl | \frac{2k}{x^2} \biggr |   
-\int_{{\tilde k}-1}^{k} dx x^{m-1}
\ln \biggl | \frac{2k}{x^2} \biggr |  \biggr ] .
\end{equation}
By carrying out the integrals on the right hand side of
Eq. (\ref{Omega-n}), we obtain, for $n=0, 1$,
\begin{eqnarray}  \label{Omega-n-fin}
\Omega_n = h_0 (k) +\sum_{m=1}^{2n+1} C_{2n+1}^m \frac{1}{m^2}(-)^m
\biggl [ (2b)^m \biggl ( m \ln \biggl | \frac{2b^2}{k} \biggr | 
-2 \biggr ) 
+ (-k)^m \biggl ( m \ln \biggl | \frac{k}{2} \biggr | 
-2 \biggr )    \nonumber \\
+({\tilde k} +1)^m \biggl ( m \ln \biggl | 
\frac{2k}{({\tilde k} +1)^2} \biggr | 
+2 \biggr ) 
+(1- {\tilde k})^m \biggl ( m \ln \biggl | 
\frac{2k}{({\tilde k} -1)^2} \biggr |
+2 \biggr )   \biggr ] .
\end{eqnarray}
One then substitutes Eqs. (\ref{Omega-1}) and (\ref{Omega-n-fin})
into Eq. (\ref{P3-App}) to obtain, with some further algebra,
Eq. (\ref{P3}). Or one can obtain first explicit expressions
for $\Omega_0$ and $\Omega_1$. In that case, one has
\begin{equation}
\Omega_0 = h_0(k) + 2 [ k \ln k -2b \ln |2b|  
+ ({\tilde k} +1) \ln |{\tilde k}+1| -({\tilde k} -1) \ln |\delta -1|],
\end{equation}
and
\begin{eqnarray}
\Omega_1 = h_0(k)+\frac{1}{3} &[& -20 k -12k \ln 2 
+ k (6+9k+2k^2) \ln k    
-2b(8 - k +2k^2) \ln |2b|   
+ ({\tilde k} +1) (11-5 {\tilde k} +2 {\tilde k}^2) \ln |{\tilde k} +1|   
\nonumber \\
&-&({\tilde k} -1) (11+5 {\tilde k} +2 {\tilde k}^2) \ln |{\tilde k} -1|].
\end{eqnarray}
\end{widetext}
In any case, the remaining derivation to obtain Eq. (\ref{P3}) 
is straightforward.

\end{document}